**LINEAR DISCRIMINANT ANALYSIS AS AN ALTERNATIVE METHOD TO INVESTIGATE THE INTERACTION OF A 1064 nm CW LASER LIGHT WITH A COLD INDUCTIVELY-COUPLED PLASMA**

M. Fatih Yılmaz[1] M. Elif Tanrıseven[2], Edgar Obonyo[3], Y.Danisman[4]

[1]Independent Researcher, Freemont, CA,USA

[2,3]Department of Education, Gazi University, Ankara, TURKEY

[4]Department of Mathematics, University of Oklahoma, Norman, OK, USA

**ABSTRACT**

In this paper, the interaction of a 1064 nm continuum-wave laser with inductively-coupled plasma generated in a fluorescent light bulb has been studied both experimentally and theoretically. The absorption coefficients pertaining to the plasma medium were obtained for different power measurements. The results indicate that absorption coefficients decrease with the increase in laser power. The UV-Vis spectra of mercury plasma were recorded by the charge-coupled spectrometer device at different power levels of laser. The linear discriminant analysis (LDA) of plasma spectra reveals the plasma ion and electron oscillations. Fourier series modeling of electron oscillation results the Whistler mode frequency of $\omega_{pe}$= 0.16 kHz with a density of $n_e$=3.9x10$^{13}$ cm$^{-3}$. 3D representation of LDA coefficients shows that the increase of laser power leads the plasma species to form in Whistler mode structures. Plasma electron temperatures ($T_e$) was inferred from the SPARTAN non-local thermal equilibrium (non-LTE) spectral code, and they were about 0.6 eV in the absence of laser light. However, there was a 0.05 eV increase in electron temperature when the laser power was absorbed by the plasma. Electron temperatures slightly increased with the increase in the power levels, which in turn resulted in smaller absorption coefficients since the absorption coefficients scales with $T_e^{-3/2}$.

## I. INTRODUCTION

Plasmas are tunable mediums, which can act as absorbers, transmitters or reflectors depending on the frequency range and application of interest. Therefore, plasmas have a widespread use in industry for many applications, such as stealth technologies in radar applications and radio communications.

Propagation of radio-frequency (RF) electromagnetic waves in uniform, non-uniform, magnetized or unmagnetized plasmas have been studied extensively both experimentally and theoretically [1]–[4]. It has been shown that by adjusting the plasma parameters, such as magnetic field strength and plasma density, it is possible to obtain high levels of absorption in a broadband range including RF and microwave frequencies [5]–[7]. However, the propagation of low-intensity laser light within cold plasmas have found less interest compared to RF or microwave electromagnetic waves emitted from antennas [4], [8]–[10]. Although laser is a special form of electromagnetic wave with a very high frequency and propagation of laser can be best described by the wave model, explanation of the absorption mechanism needs the particle model. Hence, the Drude model that explains the interaction of RF waves within plasma has failed to explain the loss mechanism of plasma when a laser beam is introduced [9]. In this context, one can apply the absorption coefficient scaling explained in [11] to interpret the loss of laser light in the plasma.

To analyze large multivariate datasets, such as plasma spectra, it is often desirable to reduce the size of the dimension of the data. Principal component analysis (PCA) and linear discriminant analysis (LDA) are the most common techniques used for this purpose. These techniques reduce the dimension of data by projecting it onto a space spanned by the vectors called principal components. These vectors are obtained successively, and they correspond to subset of data with maximum variance. The main difference between PCA and LDA is that LDA considers class labels while projecting the feature space onto a smaller space, so outperforms the PCA while discriminating the classes in database. PCA and LDA have been applied in many areas, such as medicine, robotics and remote sensing. They have also been found many applications in spectroscopy, especially in unmixing species and decomposing overlapped spectral lines of UV-VIS-NIR spectroscopy to extract the spectral fingerprints. In addition, they are used in spectroscopy of astrophysical plasmas and laboratory plasmas to extract the plasma parameters and compositions of ion species [12]–[22].

In this study, the interaction of continuum mode 1064 nm diode laser light with the inductively-coupled plasma is analyzed both experimentally and theoretically. The experimental part is based on the power-meter measurements and spectroscopy. LDA is used as a feature extractor to investigate the effects of laser power on the UV-VIS spectra of Mercury plasma. The theoretical part is based on the modeling of experimental spectra using non local thermal equilibrium (non-LTE) spectral database and laser interaction with plasma. The details of the experiment conducted are given in the next chapter, and the third chapter provides a brief summary of LDA. The fourth chapter discusses non-LTE spectral modeling and the interaction of EM-wave with the plasma. The conclusions are given in the final chapter.

## II. EXPERIMENT

The uniform magnetized plasma slab has been generated by inductively coupling a 13.56 MHz RF generator (60 Watts) to the fluorescent light bulb. The light bulb has a length of 55 cm and a diameter of 2.2 cm. Figure 1 illustrates a typical spectrum of the plasma, a schematic picture of the experiment and a view of the oscilloscope reading. The spectra have been recorded by a charge-coupled (CCD) spectrometer device of AvaSpec-ULS3648 and the images have been recorded by a CCD camera of Hero4.

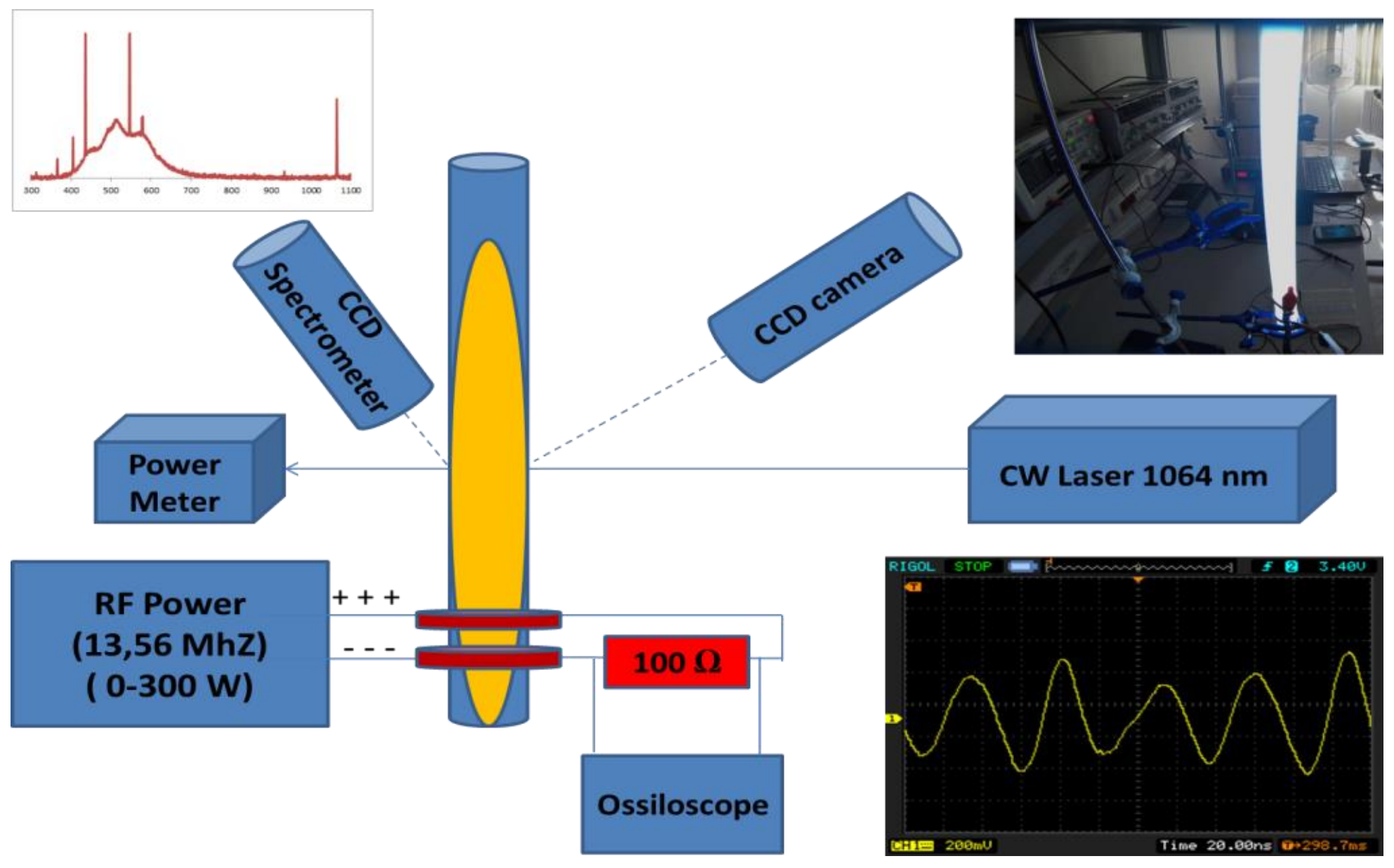

Figure 1. A schematic of the experiment and recordings.

Continuum-wave (CW) diode laser lights at different powers was directed to the plasma, and power levels were recorded by a Thorlab-PM100D power-meter. In Figure 2, applied laser powers and corresponding absorbance coefficients obtained using standard loss medium model are illustrated. Figure.2 results that low power laser attenuated quickly and the plasma becomes transparent to the laser beam as the power of laser increases. Figure 3.a illustrates the experimental spectra of mercury plasma in the absence and presence of laser light at different powers. Figure 3.b shows the zoom out of the spectral lines at 313 nm *(*$5d^{9}6s6d$*-*$5d^{9}6s6p$*)* and 365 nm ($5d^{10}6s6d$-$5d^{10}6s6p$) which are sensitive to the laser-light and slightly increases as the laser power increases.

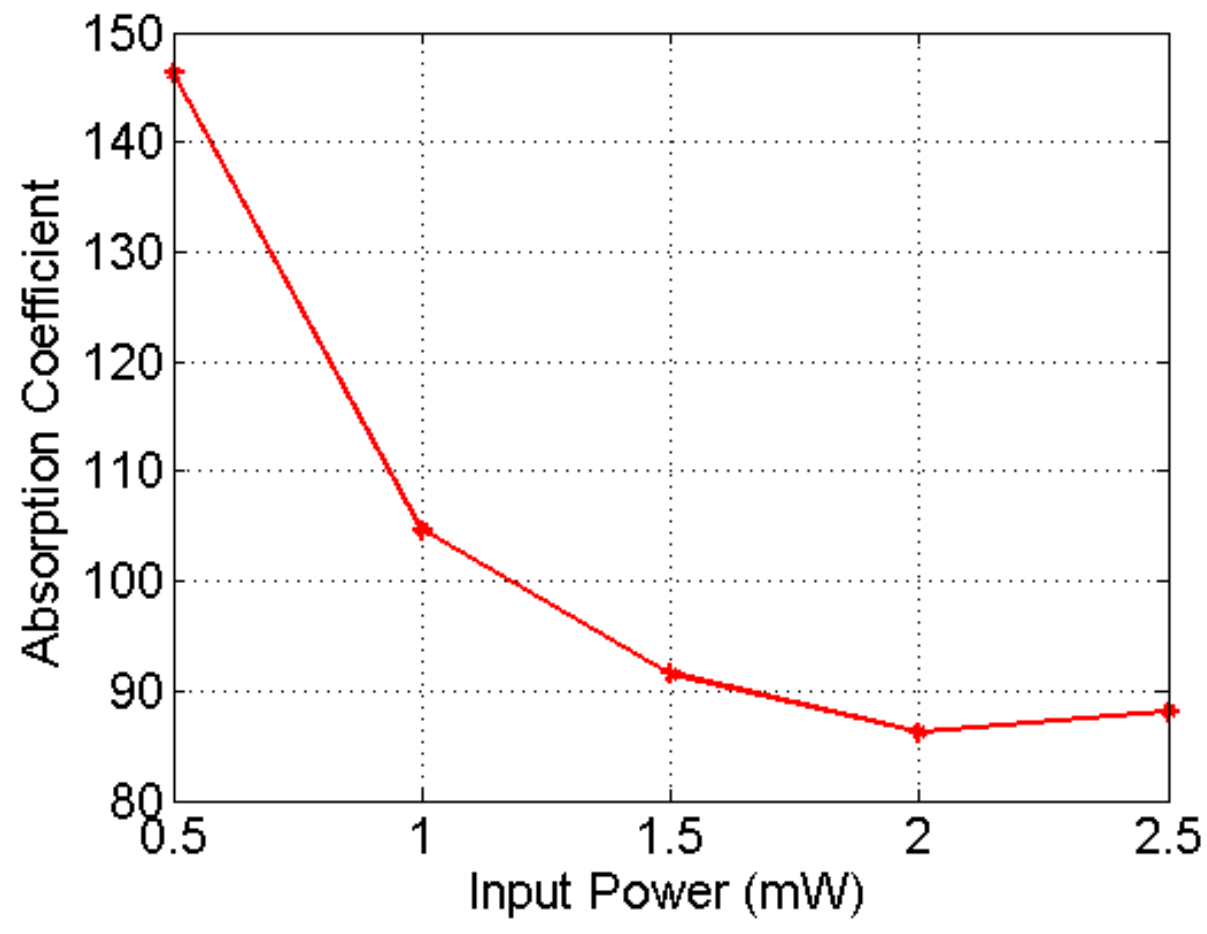

Figure 2. Absorption coefficients at different laser powers.

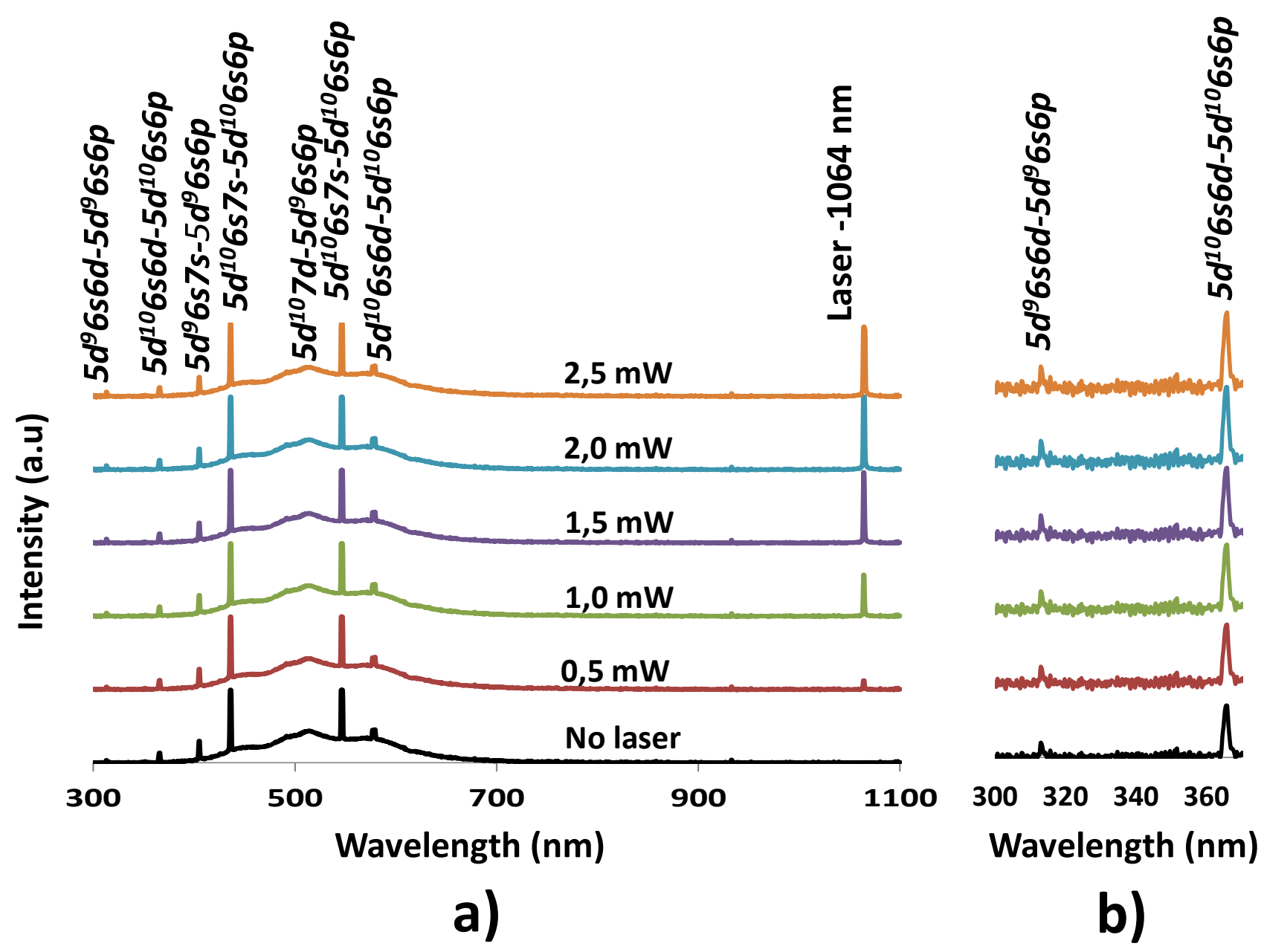

Figure 3. UV-Vis spectra of plasma in the absence and the presence of the laser light.

## III. LINEAR DISCRIMINANT ANALYSIS OF PLASMA SPECTRA

Now we provide a short summary of the mathematical background of LDA. More detailed explanation can be found in [23]. Linear Discriminant Analysis (LDA) is a dimension reduction technique to identify the hidden structures of a large data. LDA is applied to a data which consists of different classes of similar elements and used to find vectors, like principal components of PCA, to discriminate the classes while respecting the similarities among the class members. The main idea in LDA is to project a data onto a smaller subspace as in PCA but with a good class separability. In contrary, PCA deals with the entire data and does not consider the different classes. Therefore, LDA is applied to a data when different classes must be considered.

Let K be the number of classes in a dataset, and let each class consist of M elements of size N×1 . Let $\boldsymbol{\Gamma}_i^j$ be the ith element of the class j for i=1,2,...,M and j=1,2,...,K where $\boldsymbol{\Gamma}_i^j$ is a N×1 vector.

Then, within-class scatter matrix $\mathbf{S_w}$ is

$$S_w=\Sigma_{j=1}^{K} \sum_{i=1}^{M}(\Gamma_i^j - \mu_j)\,(\Gamma_i^j - \mu_j)^t \,, \tag{1}$$

where $\boldsymbol{\mu}_j$ is the mean of the class j and superscript t means the transpose, and between-class scatter matrix $\mathbf{S_b}$ is

$$S_b=\Sigma_{j=1}^{K}\,(\mu_j - \mu)(\mu_j - \mu)^t, \tag{2}$$

where $\boldsymbol{\mu}$ is the mean of all classes. In LDA, the eigenvectors of $(\mathbf{S_w})^{-1}\,\mathbf{S_b}$ provides the vectors that will be used as a basis for the new vector space. The eigenvectors corresponding to the largest two eigenvalues are called the first dominant eigenvector (|LD1>), second (|LD2>) and third dominant

eigenvector (|LD3>). As in PCA , the projection of a vector |v> onto the space spanned by |LD1> and |LD2> is

$$Proj_{< |LD1>,|LD2>,|LD3> >}|v>=w_1^l|LD1>+ w_2^l|LD2>+ w_3^l|LD3> \quad (3)$$

The coefficients $w_1^l$ , $w_2^l$ and $w_3^l$ are called the weights of |LD1>, |LD2> and |LD3> in |v> and they are equal to

$$w_1^l=|v>\blacksquare(|LD1>)^t, \quad w_2^l=|v>\blacksquare(|LD2>)^t \quad \text{and} \quad w_3^l=|v>\blacksquare(|LD3>)^t, \quad (4)$$

In this work, LDA is applied to the data obtained for the plasma spectra in the presence of laser with the input powers $P_{input}$=0.0, 0.5, 1.0, 1.5, 2.0 and 2.5 mWatt separately (Figure 3). Corresponding spectra considered for each of these cases. Hence, LDA in total is applied to 6x30=180 spectra of size 2825x1 for each case $P_{in}$=0.0, 0.5, 1.0, 1,5, 2.0 and 2.5 mWatt . Each linear discriminant component obtained is of size 2825x1.

3D plot of LD1, LD2 and LD3 coefficients which represents plasma oscillations are illustrated in Figure 4.a. As it can be seen from the figure, transitions in the absence of laser and in the presence of low laser powers follow a scattered pattern, and they finally form whistler mode structures at the laser power of 2.5 mW[24]. The plot also shows that the direction of the species changes with the increasing power levels. This is expected as the intensity of electric field associated with the laser increases with the increase in the power. |LD1> and |LD2 > spectra in Figure 4.b reveals the ion and electron oscillations, respectively [25]. Modeling of electron oscillations using Fourier series suggests the electron oscillations be around 166 Hz (0.16 kHz) which is already in the Whistler wave frequency region[26]. The plasma electron density ($n_e$=3.9x10$^{13}$ cm$^{-3}$) is obtained using the Eq.5, where $\omega_{pe}$ is the electron frequency and $\varepsilon_0$ is the permittivity of the free space

$$\omega_{pe} = \sqrt{n_e e^2 / \varepsilon_0 m_e} \quad (5)$$

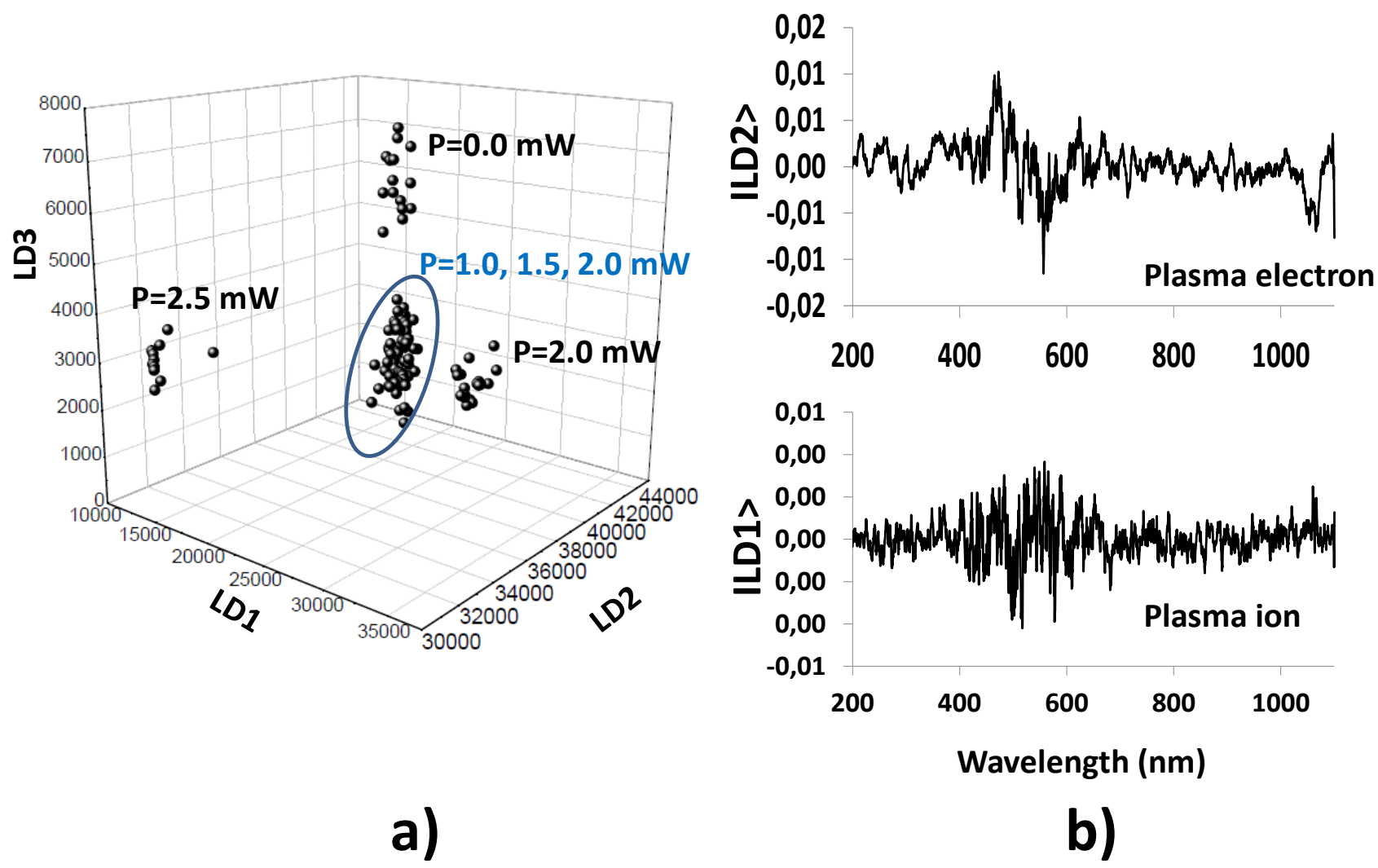

Figure 4-a) |LD1>, |LD2> and |LD3> coordinates for different laser powers and b) mean spectra, ion and electron oscillations

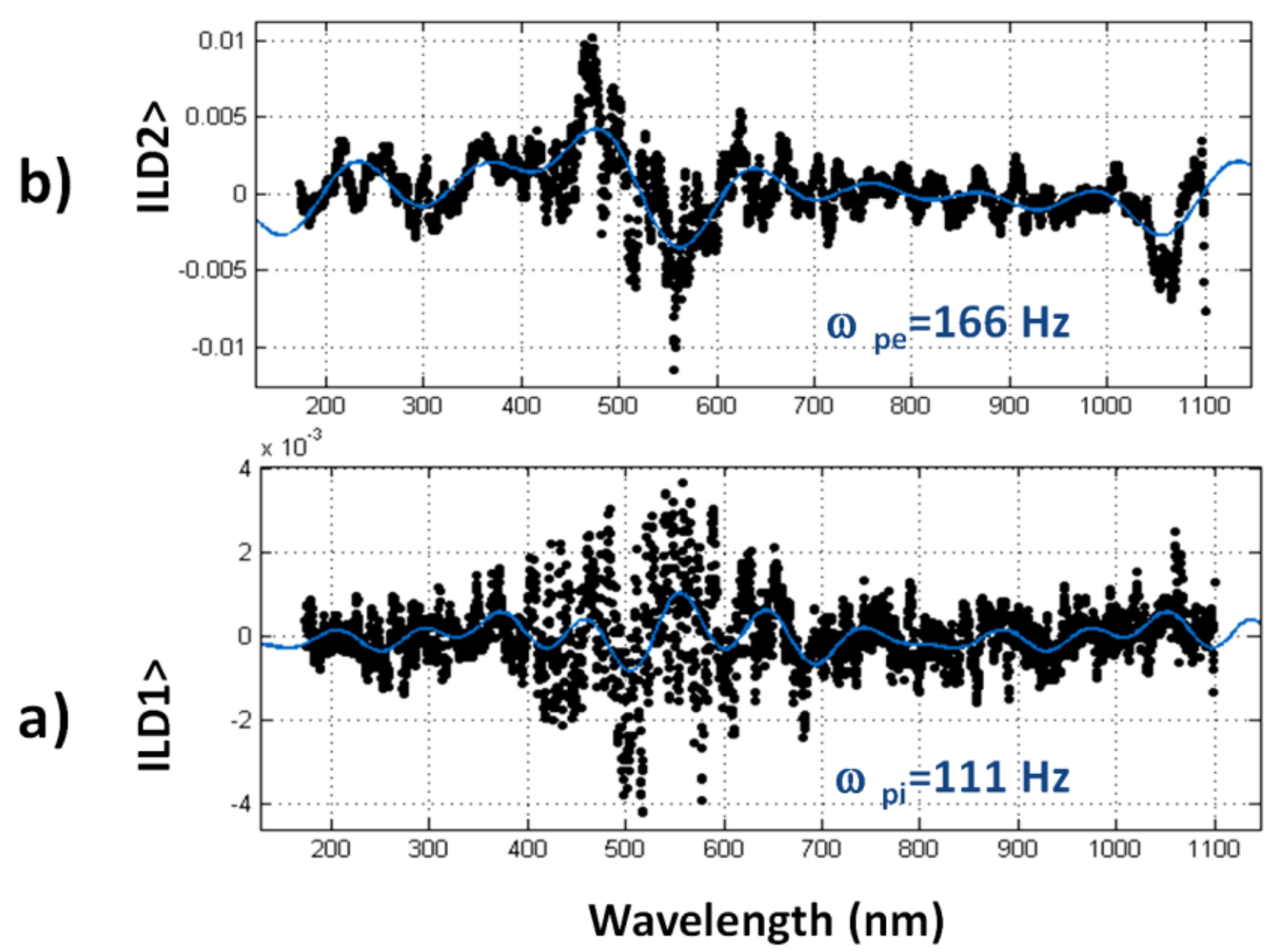

Figure 5. Fourier series fitting plasma a) ion and b) electron oscillations

## IV. A) Spectroscopic Modeling

Non thermal plasma approach can be employed in DC and RF discharge plasmas and inductively coupled plasmas (ICP). Non thermal plasmas are considered as weakly ionized low temperature plasmas. The velocity distributions of ions, and electrons follow non Maxwellian distribution [28], [29]. Electron temperatures of Mercury plasma has been estimated using the non-LTE SPARTAN code [30]. For better estimation of the plasma parameter due to scattering of laser light by the plasma, Doppler broadened profiles have been used. In Fig. 6 the temperature dependence [0.6, 0.65…, 0.85 eV] of non-LTE spectrum at electron density of $3.9 \times 10^{13}$ $cm^{-3}$ has been illustrated. Fig.5 shows that the intensity of the lines at 313 and 365 nm slightly increases as the plasma electron temperature increases. Comparison of experimental spectra and synthetic spectra suggest the plasma electron temperature be around 0.6 eV in the absence of laser light.

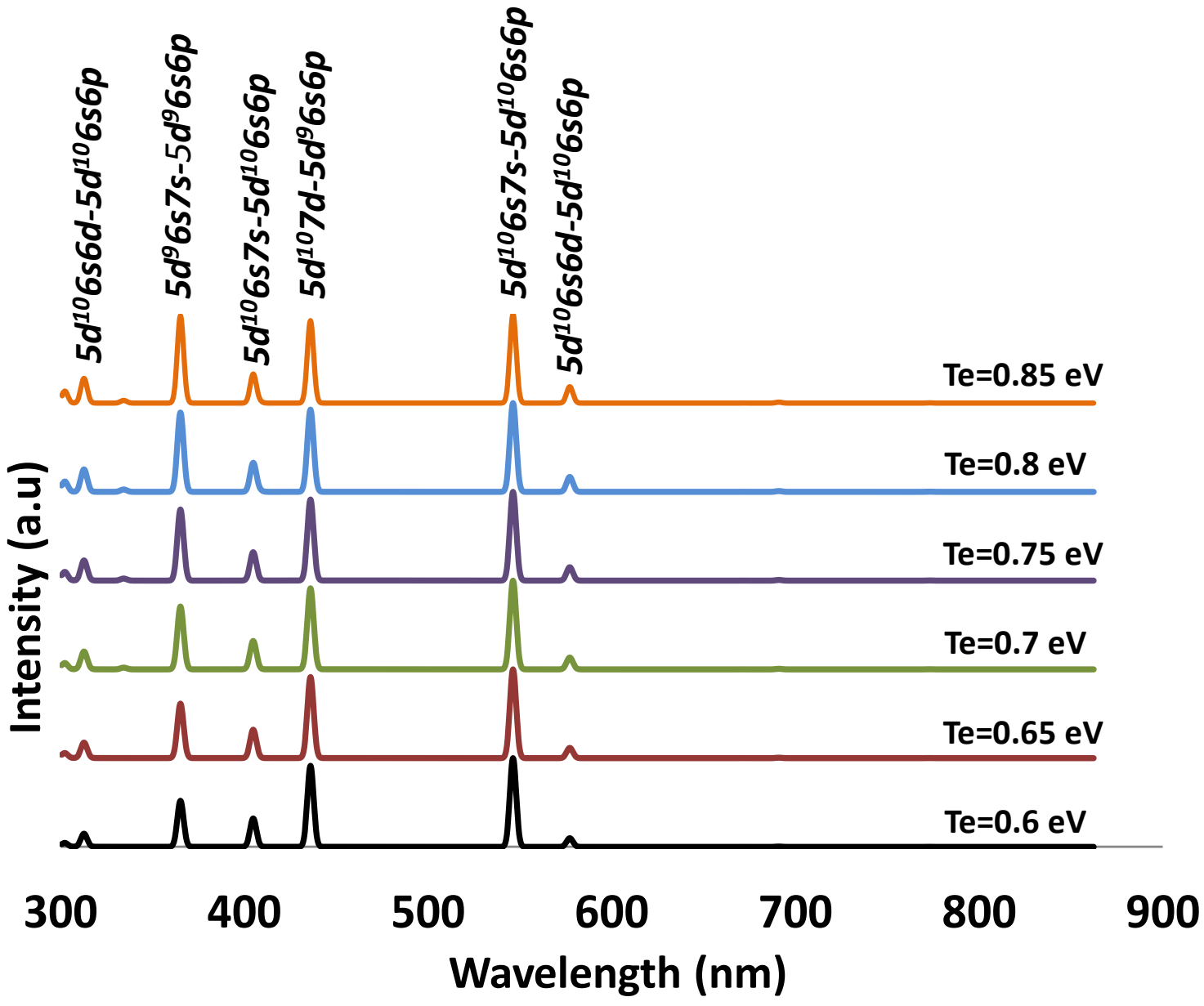

Figure 5. Plasma electron temperature dependance of non-LTE synthetic spectra at electron density of $1x\ 0^{11}cm^{-3}$

### B. Laser Plasma Interaction

For plasma mediums, the dispersion relation for electromagnetic fields can be deduced by linearizing the equation of motion for the electron fluid and coupling to the Maxwell's equations [1]. For collisionless, unmagnetized and uniform plasma, that relation is $\omega^2 = \omega_{pe}^2 + {k_c}^2 c^2$, from which the wave number for plasma can be found as

$$k = \omega \sqrt{\mu_0 \varepsilon_0} \left( 1 - \frac{\omega_{pe}^2}{\omega^2} \right)^{1/2} \qquad (6)$$

For lasers $\omega \gg \omega_p$ holds in general, and the wave number becomes almost identical to that of free space. This means that no reflection occurs at the boundaries, and laser can be transmitted through the plasma without any interaction. On the other hand, for $\omega < \omega_p$, which can hold if the plasma density is increased or the wave frequency is decreased, the wave number becomes negative and no transmission to the plasma medium occurs. However, as the laser field propagates in the plasma, the coherent motion of electrons oscillating in the laser field is converted into thermal motion by collisions with the ions in the plasma. This mechanism is called collisional or Inverse Bremsstrahlung absorption, and it introduces a loss in the system [8].

Actually, the absorption coefficients obtained by power measurements reveal that there is a significant absorption in the cold plasma. This attenuation of the laser intensity ***I*** can be modeled by

$$\frac{dI}{dx} = -\kappa_{ib} I \qquad (7)$$

where $\kappa_{ib}$ is the inverse bremsstrahlung absorption coefficient. If we denote $\rho_c$ as the mass density of the plasma corresponding to the critical density, absorption coefficient is shown to scale as

$$\kappa_{ib} \propto \left(\frac{\rho}{\rho_c}\right)^2 Z\lambda^{-2} T_e^{-3/2} \left(1 - \frac{\rho}{\rho_c}\right)^{-1/2} \qquad (8)$$

In Figure 6, we depict the absorption coefficient versus $T_e^{-3/2}$. In accordance with this scaling, an exponential increase of absorption coefficient is observed. The critical density ($n_c = 4.09\times10^{20}\,\mathrm{cm}^{-3}$) beyond which electromagnetic field propagation ceases is obtained by

$$n_c = \frac{\varepsilon_0 m_e}{e^2}\omega^2 = \frac{1.1\times10^{21}}{\lambda^2_{[\mu\mathrm{m}]}}\mathrm{cm}^{-3} \qquad (9)$$

where $\lambda = \dfrac{2\pi c}{\omega}$ is the laser wavelength in free space shows that the modeled plasma electron density is way below of the critical density.

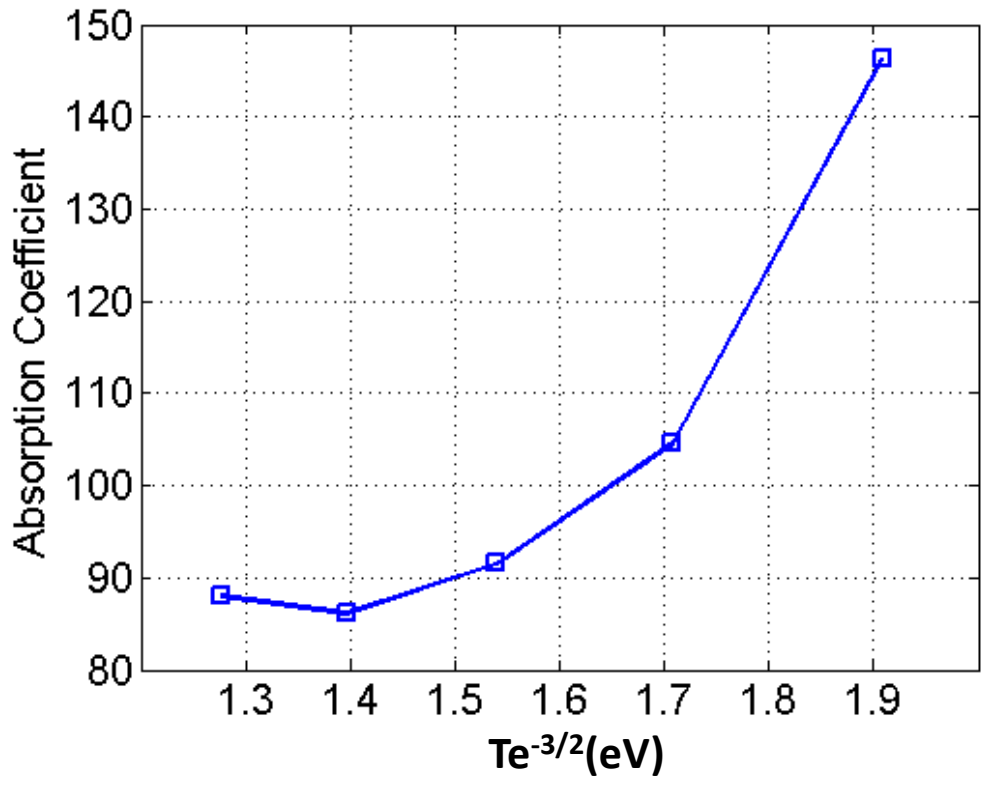

Figure 6. Absorption scaling of laser plasma interaction at different power levels.

## V. CONCLUSIONS

In this study, CW laser propagation in cold magnetized cold plasma has been investigated by means of laser power measurements and plasma emission spectroscopy. Derived absorption coefficients show that low power laser is quickly absorbed by plasma medium. LDA of experimental spectra reveals that species of plasma forms of Whistler mode structures by the increase of laser power and change their directions according to laser power. The oscillation frequency and electron density are, respectively, found to be $w_{pe}$=0.16 kHZ and $n_e=3.9\times10^{13}cm^{-3}$ by Fourier series modelling of electron oscillations. The non-LTE modeling of plasma suggests plasma electron temperature be 0.6 eV in the absence of laser light, and the temperature increases slightly by the increase of laser power. Absorption scaling based on EM-wave dispersion shows that absorbed power is converted to heat by the order of $T_e^{-3/2}$.

## ACKNOWLEDGEMENTS

This research was funded by The Scientific and Technological Research Council of Turkey (Tubitak-EEAG-113E097).

## LIST OF REFERENCES